\documentclass[twocolumn,showpacs, showkeys,nofootinbib]{revtex4}
\usepackage{amscd}
\usepackage{amsfonts}
\usepackage{amssymb}
\usepackage{amsmath}
\begin{document}
\baselineskip=11pt
\title{Possible Effects of Fierz Transformations on Vacua of Some Four-Fermion Interaction
Models \\
}
\author{Bang-Rong Zhou \footnote{E-mail:zhoubr@ucas.ac.cn}}
\affiliation{College of Physical Sciences, University of
the Chinese Academy of Sciences, Beijing 100049, China}
\date{}
\begin{abstract}
\indent A theoretical research on possible effects of the Fierz transformations on the ground
states (vacua) of some 2-flavor and $N_c$-color four-fermion (quark) interaction models has been systematically conducted. It has been shown that, based on the known criterions of the interplay between the antiquark-quark ($\bar{q}$-$q$) and diquark ($q$-$q$) condensates, in 4D space-time, for the given $\bar{q}$-$q$ channel couplings with chiral symmetry and from the heavy gluon exchange, the effects of the Fierz transformations are not enough to change the feature that the models' vacua could only be in the pure $\bar{q}$-$q$ condensate phases. However, for a given pure scalar $q$-$q$ channel coupling with the strength $H_S$, the Fierz transformations will lead to the nontrivial effect that the model's vacuum could be in the expected $q$-$q$ condensate phase only if $N_c<9$ and $H_S$ is small, and as the increase of $N_c$ and $H_S$, the vacuum will get first in a coexistence phase with $q$-$q$ and $\bar{q}$-$q$ condensates then in a pure $\bar{q}$-$q$ condensate phase until $N_c\rightarrow\infty$. Similar conclusions are also drawn from relevant four-fermion interaction models in 2D and 3D space-time. The general significance of the research is indicated.
\end{abstract}
\pacs{11.10.Lm, 12.38.Aw, 11.30.Qc, 11.15.Pg}
\keywords{four-fermion
interactions, Fierz transformations, Spinor and $U(N)$ space,
antiquark-quark and diquark channels, color number $N_c$}
\maketitle
\section{Introduction\label{Intro}}
The four-fermion interactions are very useful field theory models to
describe dynamical spontaneous breaking
\cite{kn:1,kn:2,kn:3,kn:4,kn:5,kn:6} of symmetries and their
restoring at high temperature and high density
\cite{kn:7,kn:8,kn:9,kn:10} as well as the color superconducting
phase transitions at low temperature and high density
\cite{kn:11,kn:12,kn:13}. For the involved four-fermion interaction models with dynamical symmetry breaking (from now
on the fermion will be called quark), the ground states (vacua) could be in the antiquark-quark ($\bar{q}$-$q$) condensate phase or in the diquark ($q$-$q$) condensate phase or in the coexistence phase of the above two condensates, depending on the interplay between the $\bar{q}$-$q$ and $q$-$q$ condensates in the vacua \cite{kn:14, kn:15, kn:16,kn:17, kn:18}. The presupposition of such interplay is the coexistence of the scalar $\bar{q}$-$q$ and the scalar or pseudoscalar $q$-$q$ channel couplings. On the other hand, for any given four-fermion couplings, the fermion fields entering them can always be rearranged by the Fierz transformations, thus, by the Fierz transformations, a $\bar{q}$-$q$ channel coupling will be led to some $q$-$q$ channel coupling, and the opposite case will also occur. This will inevitably lead to the coexistence of the two kinds of couplings in the resulting effective Lagrangian. Thus a natural question would be drawn out: whether the Fierz transformations could change the feature of the vacuum of a given four-fermion interaction model? A systematical research on this topic has seemly not appeared in the known literature. \\
\indent The possibility that the diquark condensates could emerge from the vacuum has been researched or touched on by some phenomenological models, including the 2 flavor Quantum Chromodynamics (QCD) instanton-induced NJL model with any $N_c$ \cite{kn:14},  the random matrix model of 2 flavor and $N_c$ color QCD \cite{kn:15} and a 2 flavor color superconducting model \cite{kn:16}. The main results show that such possibility has not been removed theoretically. To examine further this problem, we have made a more general analysis. Under the assumption that some $\bar{q}$-$q$ and $q$-$q$ channel couplings coexist, by means of the effective potential method in the mean field approximation, we have researched the interplay between the $\bar{q}$-$q$ and $q$-$q$ condensates in the vacuum respectively for 4D,2D and 3D four-fermion interaction models with flavor $N_f=2$ and color $N_c=3$ \cite{kn:17} and then extend the discussions to the case of any $N_c$ \cite{kn:18}, some useful criterions by which the $\bar{q}$-$q$ and/or $q$-$q$ condensates could emerge from the vacuum are obtained. However, in the above work, the coexistence of some $\bar{q}$-$q$ and $q$-$q$ channel couplings is only an assumption, its possible origin was not be carefully considered. Certainly, the Fierz transformations could be one of the origins, and in fact, as a check of the derived criterions, the Fierz transformations were also briefly mentioned in the Conclusions of Ref.\cite{kn:17} for some $N_c=3$ models, e.g. 4D chiral-invariant model and the heavy gluon exchange models, however, they have not constituted a systematical research on possible effects of the Fierz transformations on the vacua.\\
\indent In this paper,  we will do a systematical research on such effects. In the case of $N_f=2$ and keeping $N_c$ to be arbitrary, when some four-fermion interaction couplings are given, we will examine how their Fierz transformations  induce the couplings leading to $\bar{q}$-$q$ and $q$-$q$ condensates and how this will affect the vacuum of the model. The given starting four-fermion couplings, which in 4D case are typical and in most cases, possibly relevant to QCD-like theory, besides the chiral-invariant model and the heavy gluon exchange model, will also include the diquark channel coupling which has never been considered before. Because the strengths of the given couplings are assumed to be known, by the Fierz transformations, we will be able to fix uniquely the strengths of the $\bar{q}$-$q$ and $q$-$q$ channel couplings in the final effective Lagrangian, including their ratios. This makes it become possible, by means of the general criterions derived in Ref.\cite{kn:18}, to obtain some definite conclusions of that whether the vacua are actually in the $\bar{q}$-$q$ or $q$-$q$ condensate phase or in the coexistence phase of the two condensates.
The results will  show that in 4D space-time, for given $\bar{q}$-$q$  channel couplings, the effects of the Fierz transformations are not enough to change the models' feature that the vacua are in the pure $\bar{q}$-$q$ condensate phases. The conclusion seem a little trivial , but it is demonstrated systematically for the first time.  Furthermore,  for a given $q$-$q$ channel coupling , more  interesting non-trivial effects will emerge from the Fierz transformations. In this case, the model's vacuum could be in the expected $q$-$q$ condensate phase only if the $q$-$q$ channel coupling strength and the color number $N_c$ are small enough, otherwise, as the $q$-$q$ channel coupling strength and $N_c$ increase, the vacuum would be first in a coexistence phase with $q$-$q$  and $\bar{q}$-$q$  condensates and finally in a pure $\bar{q}$-$q$ condensate phase.  Similar conclusions will also  be derived from the 2D and 3D models. This shows some space-time dimensionality independence of the conclusions.  It is emphasized  that  the basic ideas of the above  research, including to relate the effects of the Fierz transformations to the vacua of  a class of given four-fermion interaction models with dynamical symmetry breaking, and working in the case  of any
$N_c$ and in the 4D, 2D and 3D  space-time, are original and novel, and most of the obtained results appear in the literature for the first time.\\
\indent In Sect.\ref{4D} we will analyze the effects of the Fierz transformations on scalar and pseudoscalar-isovector $\bar{q}$-$q$ channel couplings, the vectorial $\bar{q}$-$q$ channel couplings from heavy gluon exchange and  scalar $q$-$q$ channel couplings in 4D space-time and in Sect.\ref{2D} and \ref{3D}, the discussions will be extended to the similarities of the above three couplings in 2D and 3D space-time.  Finally, in Sect.\ref{conclu} we come to our conclusions. \\
\indent A brief introduction of the Fierz transformations and the explicit expressions of the Fierz transformation matrices and corresponding converse forms in spinor spaces of 4D, 2D and 3D space-time and in flavor or color $U(N)$ space will be given in Appendix. For a given $\bar{q}$-$q$ channel coupling $\mathcal{L}_{int}$, its $\bar{q}q\rightarrow \bar{q}q$ and $\bar{q}q\rightarrow qq$ channel Fierz rearrangements will be denoted respectively by $\mathcal{L}_{int}^{ex}$ (exchange terms) and $\mathcal{L}_{int}^{qq}$; For a given $qq$ channel coupling $\mathcal{L}_{qq}$, its $qq\rightarrow \bar{q}q$ channel Fierz rearrangements and corresponding exchange terms will be denoted respectively by $\mathcal{L}_{qq}^{\bar{q}q}$ and $\mathcal{L}_{qq}^{\bar{q}q-ex}$.
For a given coupling, we will put down directly the total effective Lagrangian after the Fierz transformations and focus on its physical effects. \\

\section{4D Four-fermion interactions \label{4D}}
\begin{enumerate}
  \item Scalar and pseudoscalar-isovector $\bar{q}$-$q$ channel couplings.\\
  \indent The corresponding Lagrangian may be expressed by \cite{kn:1}
\begin{eqnarray}
{\cal L}_{4(S+P\tau)}=G[(\bar{q}q)^2+(\bar{q}i\gamma_5\tau_a q)^2]
\end{eqnarray}
where $\tau_a (a=1,\cdots,N_f-1)$ are the generators of the flavor
group $SU_f(N_f)$. In present paper, the summation of a Lorentz index is implied to combine
into a Lorentz scalar and the summation of an index of the $SU(N)$
generator, unless specified otherwise, will always run over from 1
to $N^2-1$. When $N_f=2$, the above ${\cal L}_{4(S+P\tau)}$ is chiral $SU_{fL}(2)\otimes SU_{fR}(2)$-invariant \footnote{ This chiral symmetry reproduces the one of QCD with $N_c\geq 3$. Hence Eq.(1) can be related to QCD with $N_c\geq 3$. However, it can not simulate $N_c=2$ QCD with massless quarks, because the latter's chiral symmetry is the higher $SU(4)$\cite{kn:12,kn:19,kn:20,kn:21,kn:22}. It is easy to check that ${\cal L}_{4(S+P\tau)}$ in Eq.(1) does not have the $SU(4)$ symmetry. For instance, it is only a part of the whole instanton-induced four-fermion couplings which are $SU(4)$-invariant when $N_c=2$\cite{kn:21}. Based on the same grounds, the conclusions in $N_c=2$ case in present section are merely applicable for the given models here and not for $N_c=2$ QCD. In fact, the models considered here only the extensions of some $N_c=3$ QCD-relevant four-fermion interaction models to any $N_c$ case and are not supposed to touch the very special $N_c=2$ QCD theory.}.
By using the transformations (A.8),(A.17),(A.9) and (A.18) in the Appendix, we can
obtain respectively the Fierz-rearranged
 ${\cal L}_{4(S+P\tau)}^{ex}$ (exchange terms)
and ${\cal L}_{4(S+P\tau)}^{qq}$, thus the total effective Lagrangian becomes
\begin{eqnarray}
&&{\cal L}_{4(S+P\tau)}^{eff} \nonumber \\
&& ={\cal L}_{4(S+P\tau)}+{\cal L}_{4(S+P\tau)}^{ex}+{\cal
L}_{4(S+P\tau)}^{qq}
          \nonumber \\
   && =G_S(\bar{q}q)^2+G_P(\bar{q}i\gamma_5 q)^2
      +G_{P\tau}(\bar{q}i\gamma_5\tau_a q)^2\nonumber \\
   &&\hspace{0.4cm}+H_S(\bar{q}i\gamma_5\tau_A\lambda_{A^\prime}q^c)
          (\bar{q}^ci\gamma_5\tau_A\lambda_{A^\prime}q)     \nonumber \\
   &&\hspace{0.4cm}+H_P(\bar{q}\tau_A\lambda_{A^\prime}q^c)(\bar{q}^c\tau_A\lambda_{A^\prime}q)
       +\cdots
\end{eqnarray}
where we only display part of terms which could be physically
interesting, $\tau_A$ and $\lambda_{A^\prime}$ are separately the
anti-symmetric generators of the groups $SU_f(N_f)$ and $SU_c(N_c)$
and ellipsis stands for all the other possible couplings, where
$\lambda_a \,(a=1,\cdots, N_c-1)$ are the generators of $SU_c(N_c)$. It is
emphasized that ${\cal L}_{4(S+P\tau)}^{eff}$ must be used in
Hartree approximation.  When $N_f=2$, the coupling constants in Eq.(2) have the following
explicit expressions:
\begin{eqnarray}
 G_S&=&G_{P\tau}= \left(1+1/4N_c\right)G,\;\;
 G_P=-G/4N_c, \nonumber \\
 H_S&=&-H_P=G/4.
 \end{eqnarray}
Eq.(3) shows that, for the two-flavor and $N_c$-color model, the induced scalar $q$-$q$
channel interactions have a  positive coupling constant $H_S$,
however, compared with the scalar $\bar{q}$-$q$ channel interactions
with the coupling constant $G_S$, we always have the ratio
\begin{equation}
 G_S/H_S=(4N_c+1)/N_c>2/N_c.
\end{equation}
Thus, based on the general criterion of interplay between the
$\bar{q}$-$q$ and $q$-$q$ condensates \cite{kn:18}, it is impossible
to exist the scalar diquark condensates in the vacuum of this model.
Such conclusion is also valid in the limit of $N_c=2$. It is indicated that, when $N_c=2$, the scalar $\bar{q}$-$q$ condensates $\langle\bar{q}q\rangle$ and the scalar $q$-$q$ condensates $\langle\bar{q}i\gamma_5\tau_2\lambda_2q^c\rangle$ are both $SU_f(2)\otimes SU_c(2)$-singlets, however the former breaks $SU_{fL}(2)\otimes SU_{fR}(2)$ chiral symmetry but the latter conserves it. Hence in the case of $N_c=2$ we also have spontaneous breaking of the chiral symmetry.
In addition, it is noted that, after the Fierz transformations, the
largest attractive channel couplings are still the terms
$(\bar{q}q)^2$ and $(\bar{q}i\gamma_5\tau_a q)^2$ with the same
coupling strength $G_S=G_{P\tau}$, this fact certainly keeps the basic feature of the original ${\cal
L}_{4(S+P\tau)}$ in Eq.(1), including its chiral symmetry.
\item Four-fermion interactions from heavy gluon exchange.\\
\indent The corresponding Lagrangian is assumed to be \cite{kn:13}
\begin{equation}
 {\cal L}_{4(V\!\lambda)}=-g(\bar{q}\gamma^{\mu}\lambda_aq)(\bar{q}\gamma_{\mu}\lambda_aq)
\end{equation}
with the constant $g$. It simulates the interactions induced by one
gluon exchange in QCD.
Similar to the steps taken in part 1,  we can obtain the total effective
Lagrangian for $N_f=2$
\begin{eqnarray}
  {\cal L}_{4(V\!\lambda)}^{eff}&=& {\cal L}_{4(V\!\lambda)}+{\cal L}_{4(V\!\lambda)}^{ex}+{\cal L}_{4(V\!\lambda)}^{qq}
           \nonumber \\
   &=& G_S(\bar{q}q)^2+G_{V\!\lambda}(\bar{q}\gamma^{\mu}\lambda_aq)^2\nonumber \\
   &&\!\!+H_S(\bar{q}i\gamma_5\tau_A\lambda_{A^\prime}q^c)
          (\bar{q}^ci\gamma_5\tau_A\lambda_{A^\prime}q) +\cdots
  \end{eqnarray}       
  with
  \begin{eqnarray}
  G_S &=& (N_c^2-1)g/N_c^2=2(N_c-1)H_S/N_c,\nonumber\\
  G_{V\!\lambda} &=& -\left(1-1/4N_c\right)g.
\end{eqnarray}
Hence, the Fierz transformations have induced the scalar
$\bar{q}$-$q$ channel coupling and the scalar $q$-$q$ channel
coupling, however, the corresponding coupling constants $G_S$ and $H_S$,
in the case of $N_f=2$ and any $N_c$, have the ratio
\begin{equation}
 G_S/H_S=2(N_c-1)/N_c>2/N_c\,,\; \mathrm{if} \;N_c>2.
\end{equation}
This result was given in the Appendix A of Ref.\cite{kn:13}. Thus, based
on the general criterion given in
Ref.\cite{kn:18}, if $N_c\geq 3$, the ground state (vacuum) of
the model could only be in antiquark-quark condensate phase.
In the limit of $N_c=2$, $G_S/H_S=1$. This implies that we will be at a critical point between breaking and restoring of the chiral symmetry. Once there are the other couplings included, such balance would be broken and the system could come to the phase of chiral symmetry breaking or chiral symmetry restoring, depending on the feature of the included couplings.

\item Scalar diquark channel interactions.\\
\indent For describing two-flavor color superconductors, one
introduces the pure scalar $q$-$q$ channel coupling with the
Lagrangian \cite{kn:13,kn:18}
\begin{equation}
   {\cal L}_{4(S_{qq})}=H_S(\bar{q}i\gamma_5\tau_A\lambda_{A^\prime}q^c)
          (\bar{q}^ci\gamma_5\tau_A\lambda_{A^\prime}q).
\end{equation}
Eq.(9) is used usually in the case with finite quark chemical
potential, however, once it is put into a theory, then
its Fierz transformations will be bound to induce some effects even in the
case with zero quark chemical potential. In this paper we will only research
such effects on the vacuum for given pure scalar $q$-$q$ couplings.
In fact, based on the converse Fierz transformation matrices (A.10) and (A.19)
we may put down the Fierz-rearranged ${\cal
L}_{4(S_{qq})}^{\bar{q}q}$ from $q$-$q$ channel to $\bar{q}$-$q$
channel, and furthermore by using the transformation (A.8) obtain its exchange terms ${\cal
L}_{4(S_{qq})}^{\bar{q}q-ex}$ which is in fact identical to ${\cal
L}_{4(S_{qq})}^{\bar{q}q}$. Thus the effective Lagrangian for $N_f=2$ becomes
\begin{eqnarray}
  &&{\cal L}_{4(S_{qq})}^{eff}
  = {\cal L}_{4(S_{qq})}+{\cal L}_{4(S_{qq})}^{\bar{q}q}+{\cal
    L}_{4(S_{qq})}^{\bar{q}q-ex}\nonumber \\
   &&= {\cal L}_{4(S_{qq})}+G_S(\bar{q}q)^2
    +G_{P\tau}(\bar{q}i\gamma_5\tau_a q)^2\nonumber \\
       &&\hspace{0.4cm}+G_{V\!\lambda}(\bar{q}\gamma^{\mu}\lambda_{a^\prime}q)^2+\cdots,
\end{eqnarray}
\begin{eqnarray}
  G_S &=& G_{P\tau} =(N_c-1)H_S/4N_c, \nonumber \\
  G_{V\!\lambda}
  &=&-H_S/8.
\end{eqnarray}
Thus, as a result of the converse
Fierz transformations, we are led from the pure scalar $q$-$q$
channel coupling (9) to the scalar and pseudoscalar-isovector  coupling
$(\bar{q}q)^2$ and $(\bar{q}i\gamma_5\tau_a q)^2$. When $N_f=2$,
they have the same coupling strengths
$G_S=G_{P\tau}$ and this means that the chiral $SU_{fL}(2)\otimes SU_{fR}(2)$
symmetry is maintained. Meantime, we are also led to the
four-fermion interactions similar to the ones induced by heavy gluon
exchange, but with weaker strength
$G_V^{\lambda}=-H_S/8$.\\
\indent It is interesting to make a comparison between the values of
$G_S$ and $H_S$. For a given $G_S$ and $H_S$, Eq.(7) in
Ref.\cite{kn:18} has given the possible least value points of the
effective potential $V_4(\sigma, |\Delta|)$ (the ground states) of
the model with the coupling terms corresponding to $G_S$ and $H_S$,
where $\sigma$ and $\Delta$ represent the order parameters relevant
to the scalar $\bar{q}$-$q$ condensates and the scalar $q$-$q$
condensates respectively. In present case, the induced $G_S$ depends
on $H_S$ and $N_c$, and Eq.(7) in Ref.\cite{kn:18} will be reduced
to the following form: the ground state of the model will be at
\begin{widetext}
\begin{equation} (\sigma,|\Delta|)=\left\{\begin{array}{lcrll}
  (0,  \Delta_1) &\;\mbox{if}\;
  &1/2<&\tilde{H}_S<(3N_c+1)/(N_c-1)(N_c-2),\;\; &N_c<9\\
  (\sigma_2,  \Delta_2)&\;\mbox{if}\;&&\tilde{H}_S>(3N_c+1)/(N_c-1)(N_c-2),\;\;&N_c<9
  \\
  (\sigma_1,  0)&\;\mbox{if}\;&&\tilde{H}_S>4/(N_c-1), \;\;&N_c>9 \\
\end{array}\right.,
\end{equation}
\end{widetext}
where we have used the denotations $\tilde{H}_S\equiv
H_S\Lambda^2/\pi^2$ and $\Lambda$ is the 4D Euclidean momentum
cutoff of the loop integrals. It is indicated that $N_c<9$ and
$N_c>9$ correspond respectively to $G_S/H_S<2/N_c$ and
$G_S/H_S>2/N_c$. Hence, when $N_c<9$ i.e. $G_S/H_S<2/N_c$, for a
given $N_c$, the system can be in a pure $q$-$q$ condensate phase
only if the coupling strength $\tilde{H}_S$ is less than the
critical value $(3N_c+1)/(N_c-1)(N_c-2)$. Especially, in the limit of $N_c=2$, the critical
value of $\tilde{H}_S$ goes to $\infty$ and this implies that the system will only be in the chiral-invariant
pure $q$-$q$ condensate phase. Once $3\leq N_c<9$ and $\tilde{H}_S$ exceeds the above
critical value, the pure $q$-$q$ condensate phase will be
changed into a coexistence phase with the $q$-$q$ and $\bar{q}$-$q$
condensates, though the original purpose of our using ${\cal
L}_{4(S_{qq})}$ in Eq.(9) is only for expounding the pure $q$-$q$
condensates. In particular, for the realistic case with $N_c=3$ of
QCD, the expected pure $q$-$q$ condensate phase could appear only if
$\tilde{H}_S<5$. This is an interesting result. Once the strength $\tilde{H}_S$ of the given $q$-$q$ channel coupling is large enough, what could emerge from the vacuum will not be the expected diquark condensates, instead a coexistence of the $q$-$q$ and $\bar{q}$-$q$ condensates. The critical value of $\tilde{H}_S$ will decrease as
the increase of $N_c$, for example, it becomes 25/42 for $N_c=8$. \\
On the other hand, when $N_c>9$ i.e. $G_S/H_S>2/N_c$, for a
sufficiently large $\tilde{H}_S$, there could exist only the
$\bar{q}$-$q$ condensates and no the $q$-$q$ condensates. This
statement will certainly keep to be valid until
$N_c\rightarrow\infty$, consistent with the general conclusion
reached in Ref.\cite{kn:18}. The present key point is that even if
the starting point Eq.(9) is a pure scalar $q$-$q$ channel coupling,
as a result of the converse Fierz transformations, the above general
conclusion is also true.
\end{enumerate}
\section{2D Four-fermion interactions \label{2D}}
\begin{enumerate}
\item Scalar and pseudoscalar-isovector $\bar{q}$-$q$ channel couplings.\\
\indent Similar to the 4D case, we take the Lagrangian by
\begin{equation}
{\cal L}_{2(S+P\tau)}=G[(\bar{q}q)^2+(\bar{q}i\gamma_5\tau_a q)^2].
\end{equation}
However, in 2D case, we need not to consider the continuous symmetries of a Lagrangian, since they can never be spontaneously broken based on Mermin-Wagner-Coleman theorem\cite{kn:23}.
Formally Eq.(13) is the same as Eq.(1), but now the $\gamma_5$ in it
is a $2\times 2$ matrix. The steps to conduct the Fierz
transformations are similar to the ones taken in 4D
case in Sect.\ref{4D}. Based on the Fierz transformations (A.11),(A.17), (A.12) and (A.18),
the resulting total effective Lagrangian
${\cal L}_{2(S+P\tau)}^{eff}$ for $N_f=2$ becomes
\begin{equation}
{\cal L}_{2(S+P\tau)}^{eff}={\cal L}_{2(S+P\tau)}+{\cal
L}_{2(S+P\tau)}^{ex}+{\cal L}_{2(S+P\tau)}^{qq},
\end{equation}
\begin{widetext}
\begin{eqnarray}
{\cal L}_{2(S+P\tau)}^{ex}&=&\frac{G}{2N_c}\left[(\bar{q}q)^2
              +(\bar{q}i\gamma_5\tau_aq)^2
              -(\bar{q}\tau_aq)^2-(\bar{q}i\gamma_5q)^2
              -2(\bar{q}\gamma^{\mu}q)^2
              -N_c(\bar{q}\gamma^{\mu}\lambda_{a^\prime}q)^2
    \right] \nonumber  \\
  &&+\frac{G}{4}\left\{\left[(\bar{q}i\gamma_5\tau_a\lambda_{a^{\prime}}q)^2
              -(\bar{q}i\gamma_5\lambda_{a^{\prime}}q)^2\right]
              -[i\gamma_5\rightarrow 1_s]\right\},
\end{eqnarray}
\begin{eqnarray}
{\cal L}_{2(S+P\tau)}^{qq}&=&
         \frac{G}{4}\sum_{a^{\prime}=S^{\prime},A^{\prime}}\left\{\left[
              (\bar{q}i\gamma_5\tau_S\lambda_{a^{\prime}}q^c)
              (\bar{q}^ci\gamma_5\tau_S\lambda_{a^{\prime}}q)
                    -(\bar{q}i\gamma_5\tau_A\lambda_{a^{\prime}}q^c)
                     (\bar{q}^ci\gamma_5\tau_A\lambda_{a^{\prime}}q)
                       -(i\gamma_5\rightarrow 1_s)  \right] \right.\nonumber  \\
  &&\hspace{2cm}\left.-2(\bar{q}\gamma^{\mu}\tau_A\lambda_{a^{\prime}}q^c)
                        (\bar{q}^c\gamma_{\mu}\tau_A\lambda_{a^{\prime}}q)
                        \right\},
\end{eqnarray}
\end{widetext}
where $\tau_S$ and $\lambda_{S^{\prime}}$ are respectively symmetric
generators of $U_f(N_f)$ and $U_c(N_c)$, including $\tau_0\equiv
\sqrt{2/N_f}\,1_f$ and $\lambda_0\equiv\sqrt{2/N_c}\,1_c$. It it
indicated that when $N_f=2$, the coupling terms
$(\bar{q}\gamma^{\mu}\tau_aq)^2$ and
$(\bar{q}\gamma^{\mu}\tau_a\lambda_{a^{\prime}}q)^2$ have
disappeared. We see that in ${\cal L}_{2(S+P\tau)}^{eff}$ the scalar
and pseudoscalar-isovector channel couplings $(\bar{q}q)^2$ and
$(\bar{q}i\gamma_5\tau_aq)^2$ keep to be the maximal attractive
ones. Denote the respective coupling strengths by $G_S$ and $G_{P\tau}$,
then we will have
\begin{equation*}
    G_S=G_{P\tau}=\left(1+1/2N_c\right)G.
\end{equation*}
Consequently the model will maintain its original feature unchanged.
\\
\indent On the other hand, the Fierz transformations have also led
to occurrence of the scalar and pseudoscalar $q$-$q$ attractive
channel couplings $(\bar{q}i\gamma_5\tau_S\lambda_{A^{\prime}}q^c)
(\bar{q}^ci\gamma_5\tau_S\lambda_{A^{\prime}}q)$ and
$(\bar{q}\tau_A\lambda_{A^{\prime}}q^c)
(\bar{q}^c\tau_A\lambda_{A^{\prime}}q)$ with the coupling strength
$H_S=G/4$. However, considering the ratio
\begin{equation}
  G_S/H_S=2(2N_c+1)/N_c>2/N_c
\end{equation}
we can affirm similarly based on the general criterion derived in
Ref.\cite{kn:18} that for the 2D four-fermion interaction model
expressed by Eq.(13), only antiquark-quark condensates, rather than
the diquark condensates, are possible in its vacuum.
\item Four-fermion interactions from heavy gluon exchange.\\
\indent Take the Lagrangian to be
\begin{equation}
{\cal L}_{2(V\!\lambda)}=
-g(\bar{q}\gamma^{\mu}\lambda_aq)(\bar{q}\gamma_{\mu}\lambda_aq),
\end{equation}
where $\gamma^{\mu}$ are $2\times2$ matrices. After the Fierz transformations,
the total effective Lagrangian ${\cal L}_{2(V\!\lambda)}^{eff}$ for $N_f=2$
can be expressed as follows.
\begin{equation}
{\cal L}_{2(V\!\lambda)}^{eff}={\cal L}_{2(V\!\lambda)}+{\cal
L}_{2(V\!\lambda)}^{ex}+{\cal L}_{2(V\!\lambda)}^{qq}
\end{equation}
with
\begin{widetext}
\begin{eqnarray}
{\cal L}_{2(V\!\lambda)}^{ex}&=&G_S\sum_{a=0}^3
\left[(\bar{q}\tau_aq)^2+(\bar{q}i\gamma_5\tau_aq)^2\right]
      -\frac{g}{2N_c}\sum_{a=0}^3 \left[(\bar{q}\tau_a\lambda_{a^{\prime}}q)^2
            +(\bar{q}i\gamma_5\tau_a\lambda_{a^{\prime}}q)^2\right],\nonumber \\
   &&\hspace{3cm} \tau_0=1_f,\;\; G_S=(N_c^2-1)g/N_c^2
\end{eqnarray}
and
\begin{eqnarray}
{\cal
L}_{2(V\!\lambda)}^{qq}&=&-[(N_c-1)g/2N_c]\sum_{a=S,A}\left[
    (\bar{q}i\gamma_5\tau_a\lambda_{S^{\prime}}q^c)(\bar{q}^ci\gamma_5\tau_a\lambda_{S^{\prime}}q)
    +(i\gamma_5\rightarrow 1_s)\right] \nonumber \\
   &&+H_S\sum_{a=S,A}\left[
    (\bar{q}i\gamma_5\tau_a\lambda_{A^{\prime}}q^c)(\bar{q}^ci\gamma_5\tau_a\lambda_{A^{\prime}}q)
    +(i\gamma_5\rightarrow 1_s)\right], \;\;\;\;H_S=(N_c+1)g/2N_c.
\end{eqnarray}
\end{widetext}
Since $H_S>0$, so the corresponding coupling terms are attractive.
However, the ratio of the strengths of the scalar $\bar{q}$-$q$
channel coupling $(\bar{q}q)^2$ and the scalar $q$-$q$ channel
coupling $(\bar{q}i\gamma_5\tau_S\lambda_{A^{\prime}}q^c)
 (\bar{q}^ci\gamma_5\tau_S\lambda_{A^{\prime}}q)$
is obtained to be
\begin{equation}
G_S/H_S=2(N_c-1)/N_c>2/N_c,\; \mathrm{for} \;N_c>2.
\end{equation}
Hence, if $N_c\geq 3$, there will be antiquark-quark condensates alone in the
vacuum \cite{kn:18}. Eq.(22) is the same as Eq.(8) in 4D case.
\item Scalar diquark channel interactions.\\
\indent The Lagrangian is given by \cite{kn:18}\\

\begin{equation}
{\cal
L}_{2(S_{qq})}=H_S(\bar{q}i\gamma_5\tau_S\lambda_{A^\prime}q^c)
          (\bar{q}^ci\gamma_5\tau_S\lambda_{A^\prime}q).
\end{equation}

It is indicated that Eq.(23) is different from Eq.(9) with $\tau_S$
having replaced $\tau_A$ in Eq.(9), because in 2D case the matrix
$C\gamma_5$ is symmetric. We may use the converse matrices (A.13) and (A.19) in the Appendix
 to obtain
the converse Fierz-rearranged ${\cal L}_{2(S_{qq})}^{\bar{q}q}$
and furthermore use the transformation (A.11) in the Appendix to get its exchange terms
${\cal L}_{2(S_{qq})}^{\bar{q}q-ex} = {\cal L}_{2(S_{qq})}^{\bar{q}q}$,
thus the total effective Lagrangian for $N_f=2$ becomes
\begin{widetext}
\begin{eqnarray}
  {\cal L}_{2(S_{qq})}^{eff}&=&{\cal L}_{2(S_{qq})}+{\cal L}_{2(S_{qq})}^{\bar{q}q}+{\cal L}_{2(S_{qq})}^{\bar{q}q-ex}\nonumber \\
   &=& {\cal L}_{2(S_{qq})}+H_S\sum_{\Gamma^b=1_s,i\gamma_5,\gamma^{\mu}}\left[
       \frac{3(N_c-1)}{2N_c}(\bar{q}\Gamma^bq)^2
   -\frac{3}{4}(\bar{q}\Gamma^b\lambda_{a^{\prime}}q)^2
   +\frac{N_c-1}{2N_c}(\bar{q}\Gamma^b\tau_aq)^2
   -\frac{1}{4}(\bar{q}\Gamma^b\tau_a\lambda_{a^{\prime}}q)^2\right].
\end{eqnarray}
\end{widetext}
Eq.(24) contains the induced scalar channel term $(\bar{q}q)^2$
and pseudoscalar channel term $(\bar{q}i\gamma_5\tau_aq)^2$ which
respectively have the coupling strengths $G_S=3(N_c-1)H_S/2N_c$ and
$G_{P\tau}=(N_c-1)H_S/2N_c$ and it may be seen that, among all the $\bar{q}$-$q$ channel
couplings of ${\cal L}_{2(S_{qq})}^{eff}$, the scalar channel
term $(\bar{q}q)^2$ is maximal attractive. We note that the ratio of
$G_S$ and the strength $H_S$ of the scalar $q$-$q$ channel coupling
i.e. ${\cal L}_{2(S_{qq})}$ becomes
\begin{equation}
G_S/H_S =[3(N_c-1)/4](2/N_c).
\end{equation}
Based on the general criterion given in Ref.\cite{kn:18}, if there
exist the $\bar{q}$-$q$ condensates alone in the vacuum, then we
must have the condition $G_S/H_S>2/N_c$ to be satisfied, and from Eq.(25), this
implies that $3(N_c-1)/4>1$ and it leads to $N_c>7/3$. Therefore, if
$N_c\geq 3$, the vacuum of the system will in fact only in a $\bar{q}$-$q$
condensate phase, even though originally given interaction (23) is a
pure scalar $q$-$q$ channel coupling. On the other hand, if
$N_c=2<7/3$, we will have $G_S/H_S<2/N_c$, however, owing to
$G_S\neq0$, theoretically one could just acquire a mixed phase with
both $\bar{q}$-$q$ and $q$-$q$ condensates, since in 2D case, it was
proven that one could get a pure $q$-$q$ condensate phase in the vacuum only if
$G_S=0$ \cite{kn:18}.
\end{enumerate}
\section{3D Four-fermion interactions \label{3D}}
\begin{enumerate}
\item Scalar and isovector $\bar{q}$-$q$ channel couplings.\\
Since there is not $\gamma_5$ matrix in 3D space-time, the
similarities of Eqs.(1) and (13) in 4D and 2D case will be the
Lagrangian expressed by
\begin{eqnarray}
{\cal L}_{3(S+S\tau)}=G[(\bar{q}q)^2+(\bar{q}\tau_a q)^2]
\end{eqnarray}
which is $SU_c(N_c)\otimes SU_f(N_f)\otimes U_f(1)$-invariant. For
convenience, the coupling strengths of the two terms in ${\cal
L}_{3(S+S\tau)}$ are assumed to be equal, but physically this is not
essential. When $N_f=2$, by (A.14) and (A.17), the Fierz-rearranged
\begin{eqnarray*}
{\cal L}_{3(S+S\tau)}^{ex}
&=&-\frac{G}{N_c}\left[(\bar{q}q)^2+(\bar{q}\gamma^{\mu} q)^2
\right]\nonumber
\\ &&-\frac{G}{2}\left[(\bar{q}\lambda_{a^{\prime}}q)^2+(\bar{q}\gamma^{\mu}\lambda_{a^{\prime}}
q)^2\right] \hspace{1.5cm}(26a)
\end{eqnarray*}
and by (A.15) and (A.18), the Fierz-rearranged
\begin{eqnarray*}
&&{\cal L}_{3(S+S\tau)}^{qq}\\
 &&=\frac{G}{4}\sum_{a^{\prime}=S^{\prime},A^{\prime}}\left\{
    \left[(\bar{q}\tau_A\lambda_{a^{\prime}}q^c)
          (\bar{q}^c\tau_A\lambda_{a^{\prime}}q)+(1_s\rightarrow \gamma^{\mu})
          \right]\right.\nonumber \\
   &&\hspace{0.5cm}\left.-\left[(\bar{q}\tau_S\lambda_{a^{\prime}}q^c)
          (\bar{q}^c\tau_S\lambda_{a^{\prime}}q)+(1_s\rightarrow \gamma^{\mu})
          \right]\right\}.\hspace{0.9cm}(26b)
\end{eqnarray*}
 Thus the total
effective Lagrangian becomes
\begin{eqnarray}
{\cal L}_{3(S+S\tau)}^{eff} &=&{\cal L}_{3(S+S\tau)}+{\cal L}_{3(S+S\tau)}^{ex}+{\cal L}_{3(S+S\tau)}^{qq} \nonumber  \\
   &=& G_S(\bar{q}q)^2+G_{S\tau}(\bar{q}\tau_aq)^2\nonumber  \\
   && +H_P(\bar{q}\tau_A\lambda_{A^{\prime}}q^c)
          (\bar{q}^c\tau_A\lambda_{A^{\prime}}q)+\cdots,\nonumber  \\
   G_S&=&\!\!(1-1/N_c)G, \, G_{S\tau}=G,
   \,H_P=G/4.
\end{eqnarray}
It should be indicated that, after the Fierz transformations, two
maximal attractive channel couplings are still the terms
$(\bar{q}q)^2$ and $(\bar{q}\tau_aq)^2$ contained in the original
${\cal L}_{3(S+S\tau)}$. However, the two terms
with the same coupling constant $G$ in ${\cal L}_{3(S+S\tau)}$ now
have different coupling strengths $G_S<G_{S\tau}$.
This implies that  in the resulting ${\cal L}_{3(S+S\tau)}^{eff}$
the maximal attractive channel
coupling will actually be the term $(\bar{q}\tau_aq)^2$ rather than
the term $(\bar{q}q)^2$. So it is more reasonable to assume that the
condensates $\langle\bar{q}\tau_aq\rangle$ are formed more easily
than the condensates $\langle\bar{q}q\rangle$, and this will lead to
spontaneous breaking of the flavor $SU_f(N_f)$ (for $N_f=2$ i.e.
isospin) symmetry. In this case we must replace the order parameter
$\sigma=-2G_S\langle\bar{q}q\rangle$  by
$\sigma=-2G_{S\tau}\langle\bar{q}\tau_3q\rangle$ (it is possible to
fix the condensates in the $\tau_3$ direction through a rotation in
isospin space). However, it may be proven that the derived
expression for the effective potential of the model containing the
new $\sigma$ will keep unchanged in form, hence the conclusions
reached in Ref.\cite{kn:18} about interplay between the
$\bar{q}$-$q$ and $q$-$q$ condensates in the ground state (vacuum)
is still true, the mere change is to replace the scalar  channel
coupling constant $G_S$ by the scalar-isovector channel constant
$G_{S\tau}$. Since
\begin{equation}
G_{S\tau}/H_P=4>2/N_c,
\end{equation}
we can immediately conclude that although the Fierz transformations
may bring about the $q$-$q$ channel coupling corresponding to $H_P$,
it is still impossible to form the pseudoscalar diquark condensates
$\langle\bar{q}\tau_A\lambda_{A^{\prime}}q^c\rangle$ in the vacuum and
the vacuum could only in the $\langle\bar{q}\tau_3q\rangle$ condensate phase.
\item Four-fermion interactions from heavy gluon exchange.\\
\indent The Lagrangian is given by
\begin{equation}
{\cal L}_{3(V\!\lambda)}=
-g(\bar{q}\gamma^{\mu}\lambda_aq)(\bar{q}\gamma_{\mu}\lambda_aq),
\end{equation}
where $\gamma^{\mu}$ is $2\times2$ matrices in 3D space-time.\\
\indent When $N_f=2$, the Fierz-rearranged
\begin{eqnarray*}
{\cal L}_{3(V\!\lambda)}^{ex} &=&\sum_{a=0}^3
\left[G_S(\bar{q}\tau_a q)^2
-\frac{3g}{4N_c}(\bar{q}\tau_a\lambda_{a^{\prime}}q)^2 \right.\\
&& \left.-\frac{N_c^2-1}{2N_c^2}g(\bar{q}\gamma^{\mu}\tau_a q)^2
+\frac{g}{4N_c}(\bar{q}\gamma^{\mu}\tau_a\lambda_{a^{\prime}} q)^2
      \right], \\
&& G_S=3(N_c^2-1)g/2N_c^2,\hspace{2.5cm} (29a)
\end{eqnarray*}
and
\begin{eqnarray*}
{\cal L}_{3(V\!\lambda)}^{qq}&=&
H_P(\bar{q}\tau_A\lambda_{A^{\prime}}q^c)
          (\bar{q}^c\tau_A\lambda_{A^{\prime}}q)\\
          && -\frac{3(N_c-1)}{4N_c}g(\bar{q}\tau_A\lambda_{S^{\prime}}q^c)
          (\bar{q}^c\tau_A\lambda_{S^{\prime}}q) \\
          &&-\frac{N_c+1}{4N_c}g(\bar{q}\gamma^{\mu}\tau_A\lambda_{A^{\prime}}q^c)
          (\bar{q}^c\gamma_{\mu}\tau_A\lambda_{A^{\prime}}q)\\
          &&+\frac{N_c-1}{4N_c}g (\bar{q}\gamma^{\mu}\tau_A\lambda_{S^{\prime}}q^c)
          (\bar{q}^c\gamma_{\mu}\tau_A\lambda_{S^{\prime}}q)\\
          &&\hspace{1cm}+\;(\tau_A\rightarrow\tau_S),\\
          &&H_P=3(N_c+1)g/4N_c.\hspace{2.5cm}(29b)
\end{eqnarray*}
 Thus the total effective Lagrangian becomes
\begin{eqnarray*}
{\cal L}_{3(V\!\lambda)}^{eff}&=&{\cal L}_{3(V\!\lambda)}+{\cal L}_{3(V\!\lambda)}^{ex}+{\cal L}_{3(V\!\lambda)}^{qq} \\
   &=&G_S[(\bar{q}q)^2+(\bar{q}\tau_aq)^2] \\&&+H_P(\bar{q}\tau_A\lambda_{A^{\prime}}q^c)
          (\bar{q}^c\tau_A\lambda_{A^{\prime}}q) \\
   &&-\left(1-1/4N_c\right)g(\bar{q}\gamma^{\mu}\lambda_{a^{\prime}}
q)^2+\;\cdots.
\end{eqnarray*}
Since the ratio
\begin{equation}
G_S/H_P=2(N_c-1)/N_c>2/N_c, \;\;\mathrm{for}\; N_c>2\,,
\end{equation}
we can affirm that there could not be the diquark condensates in the
vacuum of the model \cite{kn:18}.
\item Pseudoscalar diquark channel coupling.\\
\indent The corresponding Lagrangian is given by
\begin{equation}
{\cal L}_{3(P_{qq})}=H_P(\bar{q}\tau_A\lambda_{A^{\prime}}q^c)
          (\bar{q}^c\tau_A\lambda_{A^{\prime}}q).
\end{equation}
By using the converse matrices (A.16) and (A.19) in the Appendix, we may obtain the Fierz-rearranged form
${\cal L}_{3(P_{qq})}^{\bar{q}q}$ of Eq.(31) from $q$-$q$ channel
to $\bar{q}$-$q$ channel, and then by Eq. (A.14) in the Appendix get its exchange terms ${\cal L}_{3(P_{qq})}^{\bar{q}q-ex} = {\cal L}_{3(P_{qq})}^{\bar{q}q}$, thus when $N_f=2$, their sum becomes
\begin{eqnarray}
 &&{\cal L}_{3(P_{qq})}^{\bar{q}q}+{\cal L}_{3(P_{qq})}^{\bar{q}q-ex}\nonumber \\&&= \frac{N_c-1}{2N_c}H_P\left[(\bar{q}\tau_aq)^2-(\bar{q}q)^2\right]\nonumber \\
   &&\;\;\;\;+\frac{H_P}{4}\left[(\bar{q}\lambda_{a^{\prime}}q)^2-(\bar{q}\tau_a\lambda_{a^{\prime}}q)^2\right] \nonumber \\
   && \;\;\;\;-\;(1_D\rightarrow\gamma^{\mu})\nonumber \\
   &&= G_{S\tau}(\bar{q}\tau_a q)^2+G_S(\bar{q}q)^2+\;\cdots,
\end{eqnarray}
where $G_{S\tau}=(N_c-1)H_P/2N_c=-G_S>0$, this implies that only the term $(\bar{q}\tau_aq)^2$ is a (maximally)
attractive interaction which could induce the isovector  condensates
$\left\langle\bar{q}\tau_aq\right\rangle$. As has been indicated in
the sector of scalar and isovector $\bar{q}$-$q$ channel couplings,
making the substitutions
$\langle\bar{q}q\rangle\rightarrow\langle\bar{q}\tau_aq\rangle$ and
$G_S\rightarrow G_{S\tau}$, we can conduct the same discussions and
reach the same conclusions as the ones obtained in Ref.\cite{kn:18,
kn:24} about interplay between the $\bar{q}$-$q$ and $q$-$q$
condensates. A special feature is now that the induced coupling
constant $G_{S\tau}$ depends on $H_P$ and $N_c$.  Let
$\sigma=-2G_{S\tau}\left\langle\bar{q}\tau_3q\right\rangle$ and
$\Delta$ represent the order parameters respectively corresponding
to the $\bar{q}$-$q$ and $q$-$q$ condensates, then the conclusion
(34) in Ref.\cite{kn:18,kn:24} will be reduced to the following
equation which shows the least value points of the effective
potential $V_3(\sigma, |\Delta|)$ being at

\begin{eqnarray}
&&(\sigma, |\Delta|)\nonumber \\
&&=\left\{
\begin{array}{clcl}
  (0, \Delta_1),&\;\;\mathrm{if}\;\; & \tilde{H}_P>1/8,   &\; N_c\leq 4<5\\
  (\sigma_1,0),&\;\;\mathrm{if} \;\;&\tilde{H}_P>1/2(N_c-1),
   &\;N_c >5\\
\end{array}\right.,\nonumber\\
\end{eqnarray}

where $\tilde{H}_P\equiv H_P\Lambda_3/\pi^2$, $\Lambda_3$ is the 3D
Euclidean momentum cutoff of the loop integrals. It is indicated
that $N_c=5$ corresponds to $G_{S\tau}/H_P=2/N_c$. We see from
Eq.(33) that the four-fermion interactions used to describe pure
pseudoscalar diquark condensates, after the converse Fierz
transformations, will induce the $\bar{q}$-$q$ channel coupling term
$(\bar{q}\tau_a q)^2$ and lead to interplay between the
$\bar{q}$-$q$ and $q$-$q$ condensates in the ground state. In this
model, the $q$-$q$ condensates could be formed only if $N_c\leq4<5$ and
in that case the ground state could be in a pure $q$-$q$
condensate phase. Once $N_c>5$, until $N_c\rightarrow\infty$,
we will be able to get merely the $\bar{q}$-$q$ condensates
$\langle\bar{q}\tau_aq\rangle$ instead of the diquark condensates.
\end{enumerate}
\section{Conclusions\label{conclu}}
In this paper, we have theoretically analyzed possible effects of the Fierz
transformations on the vacua of several given typical 4D, 2D and 3D two-flavor and
$N_c$-color four-fermion interaction models. The results can be summarized  as follows.\\
\indent It is shown that, after the Fierz transformations, the 4D
and 2D scalar and pseudoscalar-isovector couplings keep to be the
maximal attractive ones with the strength $G_S$, and some scalar diquark
channel couplings with the strength $H_S$ will be induced; in the case of 3D
scalar and isovector coupling with equal strength, the isovector
coupling will become the maximal attractive one with the strength
$G_{S\tau}$ and one also gets the induced pseudoscalar $q$-$q$
channel coupling with the strength $H_P$. However, it is found that
the resulting ratios both $G_S/H_S$ and $G_{S\tau}/H_P$ are always greater
than the the critical value $2/N_c$. This indicates that no diquark
condensates could be generated in the vacua of these models, hence the above
interaction models maintain to be the ones merely to describe possible
$\bar{q}$-$q$ condensates. The above results are valid for any $N_c$.\\
\indent For the four-fermion interactions from heavy gluon exchange,
no matter in 4D or 2D or 3D case, after the Fierz transformations,
we will always get the ratio of the induced scalar $\bar{q}$-$q$
channel coupling strength $G_S$ and the induced 4D and 2D scalar or
3D pseudoscalar $q$-$q$ channel coupling strength $H_S$ or $H_P$
expressed by
\begin{equation*}
G_S/H_S=G_S/H_P=2(N_c-1)/N_c.
\end{equation*}
On the same ground as the above, if $N_c\geq3$, this removes the possibility to
emerge the diquark condensates from the vacua and only the $\bar{q}$-$q$ condensates
could exist in the vacua.\\
\indent When the starting points are the pure diquark channel scalar
or pseudoscalar couplings with the strengths $H_S$ or $H_P$, the nontrivial effects of
the Fierz transformations on the vacua will be displayed. Owing
to the converse Fierz transformations, we will get the induced
$\bar{q}$-$q$ channel couplings including the exchange terms with the strength $G_S$ or
$G_{S\tau}$ and this is bound to lead to interplay between the
$\bar{q}$-$q$ and $q$-$q$ condensates. We have found that,
independent of dimensionality of space-time, the expected $q$-$q$ condensates
could emerge from the vacua only if $N_c<N_c^0$, a critical value determined by the
conditions $G_S/H_S<2/N_c$ or $G_S/H_P<2/N_c$ which is 9, 7/3 and
5 for 4D, 2D and 3D case respectively. \\
\indent In 4D case, when $N_c<9$, only if the coupling strength
$H_S$ is less than some $N_c$-dependent critical value, we could
just get a pure $q$-$q$ condensate phase, otherwise, will obtain a
coexistence phase with the $q$-$q$ and $\bar{q}$-$q$ condensates.
For the realistic case of $N_c=3$, the above critical value of $H_S$
corresponds to $H_S\Lambda^2/\pi^2=5$. Hence, owing to the Fierz transformations, a sufficiently
strong $q$-$q$ channel coupling could lead to not the expected  pure $q$-$q$ condensate, instead only a coexistence of the $q$-$q$ and $\bar{q}$-$q$ condensates in the vacuum. In 2D case, even
if $N_c<7/3$, we could obtain only a coexistence phase with the two condensates. In 3D case, the
condition $N_c\leq4<5$ will correspond to a pure $q$-$q$ condensate phase.\\
\indent Once $N_c>N_c^0$, until $N_c\rightarrow\infty$, in all
the cases we will obtain only the $\bar{q}$-$q$ condensates and no
the $q$-$q$ condensates in the vacua, though the original purpose to introduce
the pure $q$-$q$ channel couplings is to deal with the diquark
condensates.\\
\indent The above results show that for the models which originally do not contain the diquark channel couplings,
it seems that one needs not to worry about the occurrence of the diquark condensates in the vacua through the Fierz transformations. However, for a model of given diquark channel couplings, for example, in 4D space-time, one must note that the expected pure diquark condensates could appear only in the case of weak coupling and some small $N_c$ and this is just the nontrivial effect of the Fierz transformations on the model's vacuum to which now one must pay special attention. In addition, if one attempts to extend the above analysis based on the mean field approximation to higher order correction of the $1/N_c$ expansion, then because the above effects induced by the Fierz transformations depend on the value of $N_c$, more careful consideration must be conducted.\\
\indent The analysis made in this paper can be extended to the case of finite temperature and
finite quark chemical potential where the Fierz transformations will lead to the
interplay between the thermal $\bar{q}$-$q$ and $q$-$q$ condensates which could or could not
affect the feature of the ground state of a thermal four-fermion interaction model.\\
\indent  In this paper, we only research some special 2-flavor and $N_c$-color four-fermion models, however, the  discussions may be of more general significance for any 2-flavor and $N_c$-color four-fermion model, since for any given such model, the Fierz transformations can always lead to scalar $\bar{q}$-$q$ and scalar or pseudoscalar $q$-$q$ channel coupling and induce the interplay between the corresponding condensates in the ground state. In addition, similar or possibly different effects could be assumed to emerge from more general four-fermion interaction models with dynamical symmetry breaking. It is just the above research which first  connects the Fierz transformations with the ground states of a class of four-fermion interaction models with dynamical symmetry breaking thus provides us a new angle of view to inspect this class of models. In any case, theoretically, the possible ground state  effects of the Fierz transformations should become an implicit factor to build and treat such class of models.
\appendix*\section{Fierz
Transformations} The Fierz transformations in Dirac spinor space of
4D space-time and in $U(N)$ space can be found in Appendix A of Ref.\cite{kn:13}. However,
for this paper to be self-contained and convenience of use, we will still give a brief
introduction of the Fierz transformations and list all the necessary explicit expressions for the transformations including the new results in spinor space of 2D and 3D space-time and the converse forms of all the
non-self-converse Fierz transformations. \\
\indent Consider a local four-fermion
interactions of spinor fields $q\equiv q(x)$ with $N_f$ flavors and
$N_c$ colors, the corresponding Lagrangian is
\begin{equation}
{\cal
L}_{int}=g(\bar{q}\Gamma^aq)^2=g\Gamma_{12}^a\Gamma_{34}^a\bar{q}_1q_2\bar{q}_3q_4,
\end{equation}
where $g$ is the coupling constant, $\Gamma^a$ is outer product of
the linearly independent matrices in spinor, flavor $U_f(N_f)$ and
color $U_c(N_c)$ space, the numbers 1,2,3 and 4 represent the
indices of the elements of $\Gamma^a$, for instance, the number 1
can represent $s_1f_1c_1$ for the product matrix , or $s_1$, $f_1$
and $c_1$ when $\Gamma^a$ is separately limited to the matrix acting
on spinor, flavor and color space etc. and an index repeated always
means its summing. In view of anticommutativity of the fermion
fields $q$, Eq.(A.1) may be rewritten by
\begin{equation}
{\cal
L}_{int}=-g\Gamma_{12}^a\Gamma_{34}^a\bar{q}_1q_4\bar{q}_3q_2=:
       {\cal L}_{int}^{ex}
\end{equation}
or
\begin{equation}
{\cal L}_{int}=g\Gamma^a_{12}\Gamma^a_{34}\bar{q}_1\bar{q}_3q_4q_2=:
       {\cal L}_{int}^{qq}\,.
\end{equation}
In the above expressions, we will restrict ourselves to Hartree-type
approximation, for example, in Eq.(A.2), $\bar{q}_1$ is contracted
with $q_4$ and $\bar{q}_3$ is contracted with $q_2$ thus ${\cal
L}_{int}^{ex}$ will give exchange diagram of ${\cal L}_{int}$, and
in Eq.(A.3), $\bar{q}_1$ is contracted with $\bar{q}_3$ and $q_4$ is
contracted with $q_2$ thus ${\cal L}_{int}^{qq}$ will give the
coupling term of antiquark-antiquark ($\bar{q}$-$\bar{q}$) and
quark-quark ($q$-$q$). For this purpose, in Eq.(A.2) we must rewrite
the matrices
\begin{equation}
\Gamma^a_{12}\Gamma^a_{34}=\sum_bc_b^a\Gamma^b_{14}\Gamma^b_{32}\,,
\end{equation}
where $b$ runs over all the linearly independent matrices
$\Gamma^b$. In this paper, Eq.(A.4) will be called the Fierz
transformation of $\bar{q}$-$q\rightarrow\bar{q}$-$q$ channel. By
means of  Eq.(A.4),  Eq.(A.2) becomes
\begin{equation}
{\cal L}_{int}^{ex}=-g\sum_bc_b^a(\bar{q}\Gamma^bq)^2.
\end{equation}
On the other hand, in  Eq.(A.3) we must rewrite the matrices
\begin{equation}
\Gamma^a_{12}\Gamma^a_{34}=\sum_bd_b^a(\Gamma^bC)_{13}(C\Gamma^b)_{42},
\end{equation}
where $C$ is the charge conjugate matrix in spinor space. Eq.(A.6)
will be called the Fierz transformation of $\bar{q}$-$q\rightarrow
q$-$q$ channel. By means of Eq.(A.6), Eq.(A.3) becomes
\begin{equation}
{\cal
L}_{int}^{qq}=g\sum_bd_b^a(\bar{q}\Gamma^bq^c)(\bar{q}^c\Gamma^bq),
\end{equation}
where $q^c=C\bar{q}^T$ and $\bar{q}^c=q^TC$ are charge conjugates of
the fields $q$ and $\bar{q}$ respectively.\\
\indent The problem to solve the Fierz transformations is reduced to
calculate the expansion coefficients $c^a_b$ and $d^a_b$ in
Eqs.(A.4) and (A.6). In view of the outer product feature of
$\Gamma^a$ and the similarity of the groups $U_f(N_f)$ and
$U_c(N_c)$, we can consider separately the cases of that $\Gamma^a$
are the matrices  in spinor space and that $\Gamma^a=\{1, \tau_a\}$
with 1 as the unit matrix and $\tau_a (a=1,\cdots,N-1)$ as the
generators of the group $SU(N)$. \\

\indent In the following we will give explicit expressions of the
Fierz transformations of the matrices in spinor spaces in 4D,
2D and 3D space-time and of the $U(N)$ generators.\\
\begin{enumerate}
  \item Matrices in spinor space.
\begin{enumerate}
\item 4D space-time.\\
The independent $4\times4$ Dirac matrices are $1_s, i\gamma_5,
\gamma^{\mu}, \gamma^{\mu}\gamma_5, \sigma^{\mu\nu}\;
(\mu=0,1,2,3)$. The Fierz transformations become
\begin{widetext}
\begin{eqnarray}
 \left(
   \begin{array}{c}
     (1_s)_{12}(1_s)_{34} \vspace{0.1cm} \\
     (i\gamma_5)_{12}(i\gamma_5)_{34} \vspace{0.1cm} \\
     (\gamma^{\mu})_{12}(\gamma_{\mu})_{34} \vspace{0.1cm} \\
     (\gamma^{\mu}\gamma_5)_{12}(\gamma_{\mu}\gamma_5)_{34} \vspace{0.1cm} \\
     (\sigma^{\mu\nu})_{12}(\sigma_{\mu\nu})_{34} \\
   \end{array}
 \right)
 &=& \underbrace{\left(
        \begin{array}{rrrrr}
          \frac{1}{4} & -\frac{1}{4} & \frac{1}{4} & -\frac{1}{4} & \frac{1}{8}
          \vspace{0.1cm} \\
          -\frac{1}{4} & \frac{1}{4} & \frac{1}{4} & -\frac{1}{4} & -\frac{1}{8}
          \vspace{0.1cm} \\
          1 & 1 & -\frac{1}{2} & -\frac{1}{2} & 0 \vspace{0.1cm} \\
          -1 & -1 & -\frac{1}{2} & -\frac{1}{2} & 0 \vspace{0.1cm} \\
          3 & -3 & 0 & 0 & -\frac{1}{2} \\
        \end{array}
      \right)}_{F^{s_4}_{\bar{q}q\rightarrow\bar{q}q}}\left(
                \begin{array}{c}
                  (1_s)_{14}(1_s)_{32}  \vspace{0.1cm} \\
                  (i\gamma_5)_{14}(i\gamma_5)_{32} \vspace{0.1cm} \\
                  (\gamma^{\mu})_{14}(\gamma_{\mu})_{32} \vspace{0.1cm} \\
                  (\gamma^{\mu}\gamma_5)_{14}(\gamma_{\mu}\gamma_5)_{32}
                  \vspace{0.1cm} \\
                  (\sigma^{\mu\nu})_{14}(\sigma_{\mu\nu})_{32} \\
                \end{array}
              \right)\;\;(\bar{q}\mbox{-}q\rightarrow\bar{q}\mbox{-}q \;\mathrm{channel})
  \\
 &=&\underbrace{\left(
        \begin{array}{rrrrr}
          \frac{1}{4} & -\frac{1}{4} & \frac{1}{4} & -\frac{1}{4} & -\frac{1}{8}
          \vspace{0.1cm} \\
          -\frac{1}{4} & \frac{1}{4} & \frac{1}{4} & -\frac{1}{4} & \frac{1}{8}
          \vspace{0.1cm} \\
          1 & 1 & -\frac{1}{2} & -\frac{1}{2} & 0 \vspace{0.1cm} \\
          1 & 1 & \frac{1}{2} & \frac{1}{2} & 0 \vspace{0.1cm} \\
          -3 & 3 & 0 & 0 & -\frac{1}{2} \\
        \end{array}
      \right)}_{F^{s_4}_{\bar{q}q\rightarrow qq}}
        \left(
          \begin{array}{c}
            (i\gamma_5C)_{13}(Ci\gamma_5)_{42} \vspace{0.1cm} \\
            (C)_{13}(C)_{42}  \vspace{0.1cm} \\
            (\gamma^{\mu}\gamma_5C)_{13}(C\gamma_{\mu}\gamma_5)_{42}
            \vspace{0.1cm} \\
            (\gamma^{\mu}C)_{13}(C\gamma_{\mu})_{42}  \vspace{0.1cm} \\
            (\sigma^{\mu\nu}C)_{13}(C\sigma_{\mu\nu})_{42}  \\
          \end{array}
        \right)\;\;(\bar{q}\mbox{-}q\rightarrow q\mbox{-}q \;\mathrm{channel})
\end{eqnarray}
\end{widetext}
It is noted that the matrix $F^{s_4}_{\bar{q}q\rightarrow\bar{q}q}$
is self-converse, i.e.
$(F^{s_4}_{\bar{q}q\rightarrow\bar{q}q})^2=1_s$, thus for the
$\bar{q}$-$q\rightarrow\bar{q}$-$q$ channel, positive and converse
Fierz transformations are identical. However, the same conclusion is
not true for the $\bar{q}$-$q\rightarrow q$-$q$ channel. In fact,
the converse of the matrix $F^{s_4}_{\bar{q}q\rightarrow qq}$ is
\begin{equation}
(F^{s_4}_{\bar{q}q\rightarrow qq})^{-1} \equiv
F^{s_4}_{qq\rightarrow\bar{q}q} =\left(
        \begin{array}{rrrrr}
          \frac{1}{4} & -\frac{1}{4} & \frac{1}{4} & \frac{1}{4} & -\frac{1}{8}
          \vspace{0.1cm} \\
          -\frac{1}{4} & \frac{1}{4} & \frac{1}{4} & \frac{1}{4} & \frac{1}{8}
          \vspace{0.1cm} \\
          1 & 1 & -\frac{1}{2} & \frac{1}{2} & 0 \vspace{0.1cm} \\
          -1 & -1 & -\frac{1}{2} & \frac{1}{2} & 0 \vspace{0.1cm} \\
          -3 & 3 & 0 & 0 & -\frac{1}{2} \\
        \end{array}
      \right)
\end{equation}
which corresponds to the converse of the transformation (A.9).
Obviously, $(F^{s_4}_{\bar{q}q\rightarrow qq})^{-1}\neq
F^{s_4}_{\bar{q}q\rightarrow qq}$.

  \item 2D space-time.\\
The independent $2\times2$ matrices in spinor space are
\begin{equation*}
 1_s, \;\gamma^{\mu} (\mu=0,1),\; \gamma_5=\gamma^0\gamma^1
\end{equation*}
and the charge conjugate matrix $C=-\gamma^1$. One adoption of
$\gamma^{\mu}$ is that $\gamma^0=\sigma^3,\;\gamma^1=i\sigma^2$ with
 Pauli matrices $\sigma^i(i=1,2,3)$. The
Fierz transformations become
\begin{widetext}
\begin{eqnarray}
\left(
  \begin{array}{c}
    (1_s)_{12}(1_s)_{34} \vspace{0.1cm} \\
    (i\gamma_5)_{12}(i\gamma_5)_{34} \vspace{0.1cm} \\
    (\gamma^{\mu})_{12}(\gamma_{\mu})_{34} \\
  \end{array}
\right) &=& \underbrace{\left(
               \begin{array}{rrr}
                 \frac{1}{2} & -\frac{1}{2} & \frac{1}{2} \vspace{0.1cm} \\
                 -\frac{1}{2} & \frac{1}{2} & \frac{1}{2} \vspace{0.1cm} \\
                 1 & 1 & 0 \\
               \end{array}
             \right)}_{F^{s_2}_{\bar{q}q\rightarrow\bar{q}q}}
             \left(
               \begin{array}{c}
                 (1_s)_{14}(1_s)_{32} \vspace{0.1cm} \\
                 (i\gamma_5)_{14}(i\gamma_5)_{32}\vspace{0.1cm} \\
                 (\gamma^{\mu})_{14}(\gamma_{\mu})_{32} \\
               \end{array}
             \right)\;\;(\bar{q}\mbox{-}q\rightarrow\bar{q}\mbox{-}q \;\mathrm{channel})
                   \\
   &=&\underbrace{\left(
               \begin{array}{rrr}
                 \frac{1}{2} & -\frac{1}{2} & -\frac{1}{2} \vspace{0.1cm} \\
                 \frac{1}{2} & -\frac{1}{2} & \frac{1}{2} \vspace{0.1cm} \\
                 1 & 1 & 0 \\
               \end{array}
             \right)}_{F^{s_2}_{\bar{q}q\rightarrow qq}}
             \left(
               \begin{array}{c}
                 (i\gamma_5C)_{13}(Ci\gamma_5)_{42} \vspace{0.1cm} \\
                 (C)_{13}(C)_{42} \vspace{0.1cm} \\
                 (\gamma^{\mu}C)_{13}(C\gamma_{\mu})_{42} \\
               \end{array}
             \right)\;\;(\bar{q}\mbox{-}q\rightarrow q\mbox{-}q \;\mathrm{channel})
\end{eqnarray}
It is also indicated that
$F^{s_2}_{\bar{q}q\rightarrow\bar{q}q}=(F^{s_2}_{\bar{q}q\rightarrow\bar{q}q})^{-1}$
is self-converse; but $F^{s_2}_{\bar{q}q\rightarrow qq}$ is not, its
converse is
\begin{equation}
(F^{s_2}_{\bar{q}q\rightarrow qq})^{-1} \equiv
F^{s_2}_{qq\rightarrow\bar{q}q} = \left(
  \begin{array}{rrr}
    \frac{1}{2} & \frac{1}{2} & \frac{1}{2}  \vspace{0.1cm}\\
    -\frac{1}{2} & -\frac{1}{2} & \frac{1}{2} \vspace{0.1cm} \\
    -1 & 1 & 0 \\
  \end{array}
\right)
\end{equation}
\end{widetext}
which leads to the converse of the transformation (A.12).
\item 3D space-time.\\
\indent The independent $2\times2$ matrices in spinor space are
\begin{equation*}
    1_s,\; \gamma^{\mu} \;(\mu=0,1,2)
\end{equation*}
and the charge conjugate matrix $C=\gamma^2$, but no $\gamma_5$
exists. One adoption of $\gamma^{\mu}$ is that $\gamma^0=\sigma^3,\;
\gamma^1=i\sigma^1, \gamma^2=i\sigma^2$ with Pauli matrices
$\sigma^i (i=1,2,3)$. The Fierz transformations are
\begin{widetext}
\begin{eqnarray}
  \left(
     \begin{array}{c}
       (1_s)_{12}(1_s)_{34} \vspace{0.1cm} \\
       (\gamma^{\mu})_{12}(\gamma_{\mu})_{34} \\
     \end{array}
   \right)
   &=& \underbrace{\left(
          \begin{array}{rr}
            \frac{1}{2} & \frac{1}{2} \vspace{0.1cm} \\
            \frac{3}{2} & -\frac{1}{2} \\
          \end{array}
        \right)}_{F^{s_3}_{\bar{q}q\rightarrow\bar{q}q}}
        \left(
          \begin{array}{c}
            (1_s)_{14}(1_s)_{32}  \vspace{0.1cm} \\
            (\gamma^{\mu})_{14}(\gamma_{\mu})_{32}  \\
          \end{array}
        \right)\;\;(\bar{q}\mbox{-}q\rightarrow\bar{q}\mbox{-}q \;\mathrm{channel})
    \\
    &=& \underbrace{\left(
          \begin{array}{rr}
            -\frac{1}{2} & -\frac{1}{2} \vspace{0.1cm} \\
            \frac{3}{2} & -\frac{1}{2} \\
          \end{array}
        \right)}_{F^{s_3}_{\bar{q}q\rightarrow qq}}
        \left(
          \begin{array}{c}
            (C)_{13}(C)_{42}  \vspace{0.1cm} \\
            (\gamma^{\mu}C)_{13}(C\gamma_{\mu})_{42}  \\
          \end{array}
        \right)\;\;(\bar{q}\mbox{-}q\rightarrow q\mbox{-}q
        \;\mathrm{channel})\,.
\end{eqnarray}
Similarly, the matrix $F^{s_3}_{\bar{q}q\rightarrow\bar{q}q}$ is
self-converse, but $F^{s_3}_{\bar{q}q\rightarrow qq}$ is not. The
converse of the latter is
\begin{equation}
 (F^{s_3}_{\bar{q}q\rightarrow qq})^{-1}\equiv F^{s_3}_{qq\rightarrow \bar{q}q} =
 \left(
   \begin{array}{rr}
     -\frac{1}{2} & \frac{1}{2} \vspace{0.1cm} \\
     -\frac{3}{2} & -\frac{1}{2} \\
   \end{array}
 \right)
\end{equation}
\end{widetext}
which will generate the converse of the transformation (A.15).
\end{enumerate}
\item Generators of $U(N)$.\\
\indent We denote the generators of the group $SU(N)$ by $\tau_a
(a=1,\cdots,N-1)$ and define $\tau_0\equiv\sqrt{2/N}\,1$, where 1 is
the $N\times N$ unit matrix, they are normalized by
$\mathrm{Tr}\,\tau_a\tau_b=2\delta_{ab}$. \\
\indent The Fierz transformations may be expressed by
\begin{widetext}
\begin{eqnarray}
    \left(
       \begin{array}{c}
         (1)_{12}(1)_{34} \vspace{0.1cm} \\
         (\tau_a)_{12}(\tau_a)_{34} \\
       \end{array}
     \right)&=&\underbrace{\left(
                    \begin{array}{cc}
                      \frac{1}{N} & \frac{1}{2} \vspace{0.1cm} \\
                      2\frac{N^2-1}{N^2} & \frac{1}{N} \\
                    \end{array}
                  \right)}_{F^{U(N)}_{\bar{q}q\rightarrow\bar{q}q}}
                 \left(
                   \begin{array}{c}
                     (1)_{14}(1)_{32} \vspace{0.1cm} \\
                     (\tau_a)_{14}(\tau_a)_{32} \\
                   \end{array}
                 \right)\;\;(\bar{q}\mbox{-}q\rightarrow\bar{q}\mbox{-}q \;\mathrm{channel})
            \\
          &=&\underbrace{\left(
                    \begin{array}{cc}
                      \frac{1}{2} & \frac{1}{2}\vspace{0.1cm} \\
                      \frac{N-1}{N} & -\frac{N+1}{N} \\
                    \end{array}
                  \right)}_{F^{U(N)}_{\bar{q}q\rightarrow qq}}
                 \left(
                   \begin{array}{c}
                     (\tau_S)_{13}(\tau_S)_{42} \vspace{0.1cm} \\
                     (\tau_A)_{13}(\tau_A)_{42} \\
                   \end{array}
                 \right)\;\;(\bar{q}\mbox{-}q\rightarrow q\mbox{-}q \;\mathrm{channel})
\end{eqnarray}
\end{widetext}
In Eq.(18) $\tau_S$ (including $\tau_0$) and $\tau_A$ are respectively symmetric and anti-symmetric generators of $U(N)$. It is indicated that the matrix
$F^{U(N)}_{\bar{q}q\rightarrow\bar{q}q}$ is self-converse, but
$F^{U(N)}_{\bar{q}q\rightarrow qq}$ is not. The converse of the
latter is
\begin{equation}
 (F^{U(N)}_{\bar{q}q\rightarrow qq})^{-1}\equiv F^{U(N)}_{qq\rightarrow \bar{q}q} =
 \left(
   \begin{array}{rr}
     \frac{N+1}{N} & \frac{1}{2} \vspace{0.1cm} \\
     \frac{N-1}{N} & -\frac{1}{2} \\
   \end{array}
 \right)
\end{equation}
which corresponds to the $U(N)$ Fierz transformation of
$qq\rightarrow\bar{q}q$ channel. \\
\indent It is emphasized that in the discussions of this paper, when
$U(N)$ is considered as the flavor group $U_f(N_f)$, the generators
will be denoted by $(1_f, \tau_a)$ or $(\tau_S, \tau_A)$ and when
$U(N)$ is considered as the color group $U_c(N_c)$, the generators
will be denoted by $(1_c, \lambda_a)$ or $(\lambda_S, \lambda_A)$.
\end{enumerate}

\end{document}